\documentclass[prd,amsmath,amssymb,floatfix,superscriptaddress,nofootinbib,twocolumn]{revtex4-1}

\def\be{\begin{equation}}
\def\ee{\end{equation}}
\def\bea{\begin{eqnarray}}
\def\eea{\end{eqnarray}}
\newcommand{\vs}{\nonumber\\}

\newcommand{\vk}{\vec{k}}
\newcommand{\ec}[1]{Eq.~(\ref{eq:#1})}

\newcommand{\eql}[1]{\label{eq:#1}}
\newcommand{\rf}[1]{\ref{fig:#1}}
\newcommand{\sfig}[2]{
\includegraphics[width=#2]{#1}
        }

\newcommand{\sfiggg}[2]{
\includegraphics[width=0.8\paperwidth]{#1}
        }

\newcommand{\Sfig}[2]{
   \begin{figure}[thbp]
   \begin{center}
    \sfig{#1.pdf}{\columnwidth}
    \caption{{\small #2}}
    \label{fig:#1}
     \end{center}
   \end{figure}
}
\newcommand{\Sfigg}[2]{
   \begin{figure}[thbp]
    \sfiggg{#1.pdf}{.8\paperwidth}
    \caption{{\small #2}}
    \label{fig:#1}
   \end{figure}
}

\usepackage[utf8]{inputenc}
\usepackage{graphicx}
\usepackage{amssymb}
\usepackage{simplewick}
\usepackage{amsmath}
\usepackage{bm}
\usepackage{color}
\usepackage{enumitem}
\usepackage[linktocpage=true]{hyperref} 
\hypersetup{
    colorlinks=true,       
    linkcolor=red,         
    citecolor=blue,        
    filecolor=magenta,      
    urlcolor=blue           
}
\usepackage[all]{hypcap} 
\definecolor{darkgreen}{cmyk}{0.85,0.1,1.00,0} 
\definecolor{darkorange}{rgb}{1.0,0.2,0.0}

\begin{document}
\title{Large Scale Structure Reconstruction with Short-Wavelength Modes}
\author{\large Peikai Li}
\affiliation{Department of Physics, Carnegie Mellon University, Pittsburgh, PA 15213, USA}
\affiliation{McWilliams Center for Cosmology, Carnegie Mellon University, Pittsburgh, PA 15213, USA}
\author{\large Scott Dodelson}
\affiliation{Department of Physics, Carnegie Mellon University, Pittsburgh, PA 15213, USA}
\affiliation{McWilliams Center for Cosmology, Carnegie Mellon University, Pittsburgh, PA 15213, USA}
\author{\large Rupert A. C. Croft}
\affiliation{Department of Physics, Carnegie Mellon University, Pittsburgh, PA 15213, USA}
\affiliation{McWilliams Center for Cosmology, Carnegie Mellon University, Pittsburgh, PA 15213, USA}

\date{\today}
\begin{abstract}
\noindent Large scale density modes are difficult to measure because they are sensitive to systematic observational errors in galaxy surveys, but we can study them indirectly by observing their impact on small scale perturbations. Cosmological perturbation theory predicts that second-order density inhomogeneities are a convolution of a short- and a long-wavelength mode. This arises physically because small scale structures grow at different rates depending on the large scale environment in which they reside. This induces an off-diagonal term in the two-point statistics in Fourier space that we use as the basis for a quadratic estimator for the large scale field. We demonstrate that this quadratic estimator works well on an N-body simulation of size  $(2.5\,h^{-1}\,\rm Gpc)^3$. In particular, the quadratic estimator successfully reconstructs the long-wavelength modes using only small-scale information. This opens up novel opportunities to study structure on the largest observable scales. 

\end{abstract}
\maketitle
\section{Introduction}
Measuring the distribution of matter on large scales is one of the goals of cosmological surveys \cite{LSST:2012ls}\cite{Wfirst:2012jg}. The information contained on large scales may provide information about issues ranging from the turnover in the power spectrum (and therefore the total matter density) to the accelerated universe to anomalies observed in the cosmic microwave background to primordial nongaussianity. In \cite{Karkare:2019delen}, a method of using delensing with intensity mapping has been proposed to directly measure large scale modes. But generally speaking, direct measurements are difficult because of observational systematic effects, so indirect approaches have been considered. As pointed out by \cite{Modi:2019hydr}, 21cm intensity mapping is one area where due to foregrounds, large-scale (line of sight) modes will be impossible to measure directly.

Small scale structure grows differently in the presence of an large-scale overdensity: it is as if the mean background density is larger than on average. This relation between long- and short-wavelength modes has been discussed in recent years \cite{Masui:2010prim}\cite{Baldauf:2011fer}\cite{Jeong:2012foss}\cite{Jasche:2013bay}\cite{Leclercq:2013remp}\cite{Li:2014ssc}\cite{Zhu:2016tidal}\cite{Barreira:2017res}\cite{Seljak:2017opt}\cite{Akitsu:2018dist}\cite{Foreman:2018line}\cite{Zhu:2018cm}. The method of using the small scale position-dependent power spectrum to compute the squeezed-limit bispectrum also indicates that small scale perturbations can be used to infer large scale information \cite{Chiang:2014pos}\cite{Chiang:2015poss}. In order to fully achieve this goal, here we construct a quadratic estimator to measure long-wavelength modes indirectly. 

Standard perturbation theory (SPT) \cite{Goroff:1986sts}\cite{Makino:1992fs}\cite{Jain:1994sop} identifies the second-order contribution to a short-wavelength mode as a convolution of a short- and a long-wavelength mode. Abstractly, this is similar to
cosmic microwave background (CMB) lensing~\cite{Hu:2001dt}\cite{Hu:2002mr}, where the CMB temperature field has a second-order correction due to the gravitational field along the line of sight. Similarly in our case the short-wavelength mode's nonlinear terms are related to its large scale environment.
The construction of a CMB lensing quadratic estimator makes use of the fact that small scale two-point correlations of CMB temperature modes have off-diagonal terms due to large scale perturbations caused by gravitational lensing. The same statistical feature shows up in our case as well -- the off-diagonal terms of the small scale correlations are no longer zero, due to the effect of large scale modes. Thus we can create a quadratic estimator for long-wavelength modes using exactly the same formalism. 

We begin with a brief review of SPT up to second-order, build the quadratic estimator, and then assess its detectability. We then apply the estimator to data from a large N-body simulation and demonstrate that it successfully extracts the large scale modes. 
We use a flat $\Lambda$CDM model with Planck Collaboration XVI (2014)~\cite{Planck:2014cos} cosmological parameters in this work (to match the parameters of the N-body simulation).

\section{Standard Perturbation Theory}\label{sec1}
Starting from a perfect pressureless fluid, the nonrelativistic cosmological fluid equations are the continuity, Euler and Poisson equations:
\bea
\frac{\partial \delta(\vec{x},\tau)}{\partial \tau} &+&\vec{\nabla}\cdot [(1+\delta(\vec{x},\tau))\vec{v}(\vec{x},\tau)] =0\eql{fluid1} \\
\bigg[\frac{\partial}{\partial \tau} +\vec{v}(\vec{x},\tau)\cdot&\vec{\nabla}& \bigg]\vec{v}(\vec{x},\tau)=-\frac{da}{d\tau}\frac{\vec{v}(\vec{x},\tau)}{a}-\vec{\nabla}\Phi \eql{eulereq}\\
&&\nabla^2 \Phi = 4\pi G a^2 \bar{\rho}_{\rm m} \delta(\vec{x},\tau) .\eql{fluid3}
\eea
Here $a$ is the cosmological scale factor, $\Phi$ is the 3D gravitational potential and $\bar{\rho}_{\rm m}(a)$ is the mean matter density. These equations fully determine the time evolution of the local density contrast $\delta$ and the peculiar velocity field $\vec{v}=d\vec{x}/d\tau$.
We can solve these equations perturbatively in Fourier space~\cite{Bernardeau:2002rev}:
\bea
{\delta}(\vk,\tau) &=&\sum_{n=1}^{\infty} {\delta}^{(n)}(\vk,\tau)=\sum_{n=1}^{\infty}D_1^{n}(\tau)\delta_{n}(\vk)  \eql{pert1}\\
{\theta}(\vk,\tau)&=&\sum_{n=1}^{\infty}{\theta}^{(n)}(\vk,\tau) \vs
&=&-\frac{d\,\ln D_1(\tau)}{d\tau}\sum_{n=1}^{\infty}D_1^{n}(\tau)\theta_{n}(\vk) \eql{pert2}
\eea
where $D_1$ is the linear growth factor. 
The first order term ${\delta}^{(1)}$ corresponds to linear evolution. The linear power spectrum is given by this first order term via averaging over modes in Fourier space:
\be 
\langle {\delta^{(1)}}(\vk,\tau){\delta^{(1)}}(\vk',\tau) \rangle =(2\pi)^3 \delta_{\rm D}(\vk+\vk')P_{\rm lin}(k,\tau) \eql{lin}
\ee 
Here $\delta_{\rm D}$ is the Dirac delta function. Substituting the perturbative series \ec{pert1} and \ec{pert2} into the Fourier transformed fluid equations \ec{fluid1}-\ec{fluid3} leads to an expression for the second-order density contrast:
\be
{\delta}^{(2)}(\vk,\tau)=\int \frac{d^{3}\vk_{1}}{(2\pi)^3} F_2(\vk_1,\vk-\vk_1){\delta}^{(1)}(\vk_1,\tau) {\delta}^{(1)}(\vk-\vk_1,\tau) \eql{sorder} 
\ee
with 
\be
F_{2}(\vk_1,\vk_2)=\frac{5}{7}+\frac{2}{7}\frac{(\vk_1\cdot \vk_2)^2}{k_1^2 k_2^2}+\frac{\vk_1\cdot \vk_2}{2k_1k_2}\bigg[\frac{k_1}{k_2}+\frac{k_2}{k_1}\bigg].\eql{f2}
\ee
Note that \ec{pert1}, \ec{pert2} and \ec{f2} are completely accurate only in an Einstein-de Sitter universe and also assuming the case of a pressureless perfect fluid. Nonetheless for related calculations in a $\Lambda$CDM universe, the difference is found to be negligible \cite{Takahashi:2008to}, and thus we use the expressions from \ec{pert1} and \ec{f2} throughout this work.
Using this expression for $\delta^{(2)}$, we can calculate the two-point correlation of two short-wavelength modes $\vk_s$ and $\vk_s'$, in the squeezed limit $\vk_l=\vk_s+\vk_s'$ with $\vk_s,\vk_s' \gg \vk_l$. Here $\vk_l$ corresponds to a long-wavelength mode, and we suppress the time dependence. To second order,
\bea 
 \langle {\delta}(\vec{k}_s){\delta}(\vec{k}_s') \rangle|_{\vk_s+\vk_s'=\vk_l}&=&
  \langle {\delta}^{(1)}(\vec{k}_s){\delta}^{(2)}(\vec{k}_s') \rangle\vs
  &+&\langle {\delta}^{(2)}(\vec{k}_s){\delta}^{(1)}(\vec{k}_s') \rangle.
\eea 
Substituting \ec{sorder} into the first bracket we get:
\bea 
\langle {\delta}^{(1)}(\vec{k}_s){\delta}^{(2)}(\vec{k}_s') \rangle =  \int \frac{d^3\vec{k}}{(2\pi)^3} F_2 (\vec{k},\vec{k}_s'-\vec{k})\vs
\times \langle {\delta}^{(1)}(\vec{k}_s){\delta}^{(1)}(\vec{k}_s'-\vec{k}){\delta}^{(1)}(\vec{k}) \rangle \eql{integral}
\eea 
When one of the wavenumbers in the 3-point function in \ec{integral} is very small, that mode can be considered as a background mode. The small scale modes evolve in the presence of whatever long wavelength modes happen to be present. Therefore,  we can take the long-wavelength mode out of the bracket:
\bea
\langle
\delta^{(1)}(\vk_{s}) \delta^{(1)}(\vk_{s}'-\vk){\delta^{(1)}}(\vk) 
\rangle
&=&\langle {\delta^{(1)}}(\vk_{s}) {\delta^{(1)}}(\vk_{s}'-\vk) \rangle {\delta^{(1)}}(\vk)  \vs
&+&\langle {\delta^{(1)}}(\vk_{s}) {\delta^{(1)}}(\vk) \rangle {\delta^{(1)}}(\vk_s'-\vk) .
\vs
\eql{contraction}
\eea
\Sfig{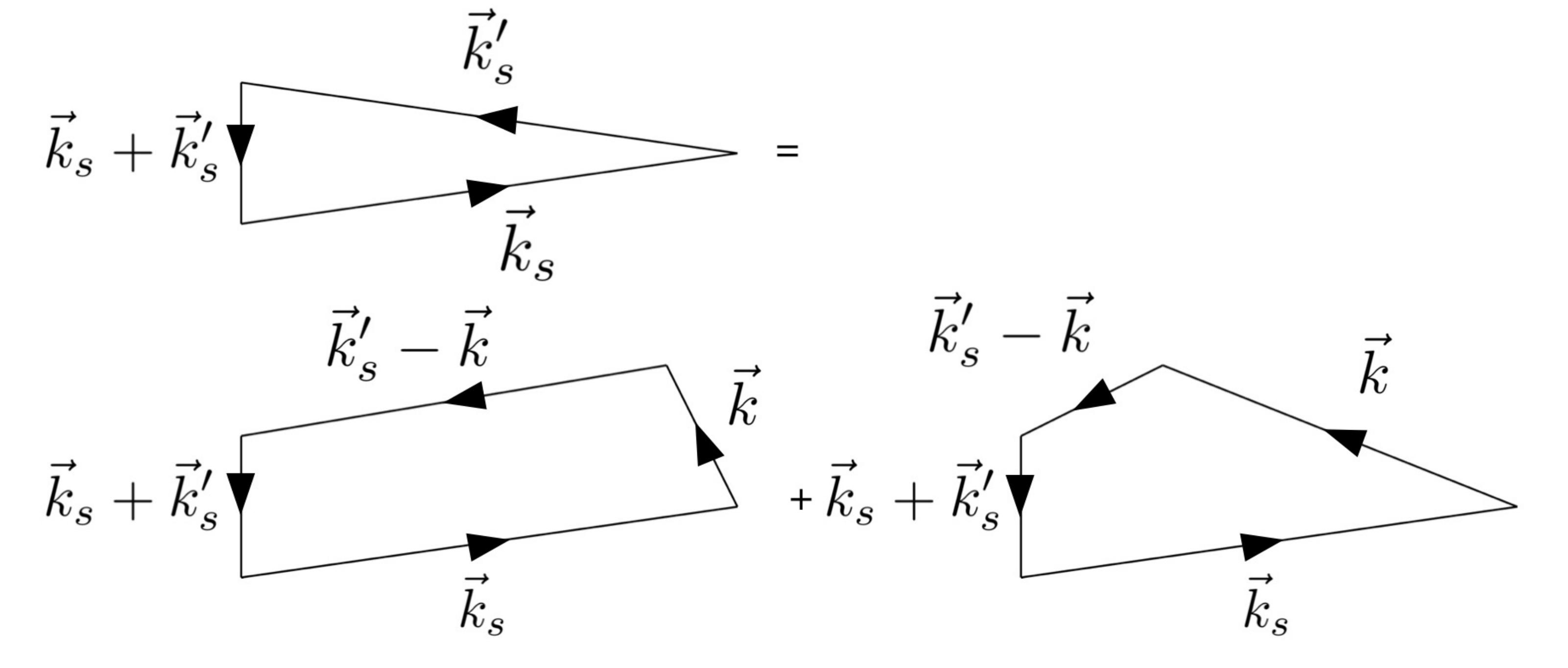}{\ec{contraction} occurs when $\vec{k}\ll\vec{k}_s,\vec{k}'_s$ or $\vec{k}'_s-\vec{k}\ll\vec{k}_s,\vec{k}'_s$, which corresponds to the two terms shown in this figure.}
As shown in Fig.~\rf{LWM}, the first term on the right occurs when $\vk$ is small and the second when $\vk_s'-\vk$ is small.
Using \ec{lin}, \ec{integral} then becomes:
\bea 
&&\int \frac{d^3\vec{k}}{(2\pi)^3} F_2 (\vec{k},\vec{k}_s'-\vec{k})\langle {\delta}^{(1)}(\vec{k}_s){\delta}^{(1)}(\vec{k}_s'-\vec{k}){\delta}^{(1)}(\vec{k}) \rangle \vs
&=&\int d^3\vk F_2(\vec{k},\vec{k}_s'-\vec{k}) \delta_{\rm D}(\vk_s+\vk_s'-\vk)P_{\rm lin}(k_s){\delta}^{(1)}(\vec{k}) \vs
&+&\int d^3\vk F_2(\vec{k},\vec{k}_s'-\vec{k}) \delta_{\rm D}(\vk_s+\vk)P_{\rm lin}(k_s){\delta}^{(1)}(\vk_s'-\vk) \vs
&=&2F_2(-\vk_s,\vk_s+\vk_s')P_{\rm lin}(k_s){\delta}^{(1)}(\vk_s+\vk_s')
\eea 
where we take advantage of the fact that $F_2$ is a symmetric function. Finally we have:
\be 
\langle {\delta}(\vec{k}_s){\delta}(\vec{k}_s') \rangle =f(\vec{k}_s,\vec{k}_s'){\delta}^{(1)}(\vec{k}_l) \eql{2pt}
\ee 
with
\bea
f(\vec{k}_s,\vec{k}_s')&=&2F_2(-\vec{k}_s,\vec{k}_s+\vec{k}_s')P_{\rm lin}(k_s)\vs
&+&2F_2(-\vec{k}_s',\vec{k}_s+\vec{k}_s')P_{\rm lin}(k_s')       
\eea 
This suggests that we can estimate long-wavelength modes using short-wavelength modes. Since the left-hand side of \ec{2pt} has only short modes while the right-hand side of it is sensitive to long modes.

\section{Quadratic Estimator}\label{sec2}
We can now construct the quadratic estimator for long-wavelength modes starting from \ec{2pt} and summing over as many pairs as possible with weights that maximize the signal-to-noise. As with the case of CMB lensing, we can write the general form of the estimator by averaging over pairs of short-wavelength modes:
\begin{eqnarray}
\hat{\delta}^{(1)}(\vec{k}_l)=A(\vec{k}_l)\int \frac{d^3 \vec{k}_s}{(2\pi)^3} g(\vec{k}_s,\vec{k}_s'){\delta}(\vec{k}_s){\delta}(\vec{k}_s') \eql{quadest},
\end{eqnarray} 
with $g$ being a weighting function, $\vk_s'=\vk_l-\vk_s$ and $A$ is the normalization prefactor defined by requiring that $\langle \hat{\delta}^{(1)}(\vec{k}_l) \rangle={\delta}^{(1)}(\vec{k}_l)$:
\begin{eqnarray}
A(\vec{k}_l)=\bigg[\int \frac{d^3 \vec{k}_s}{(2\pi)^3} g(\vec{k}_s,\vec{k}_s')f(\vec{k}_s,\vec{k}_s')  \bigg]^{-1} \eql{a}
\end{eqnarray}
In the absence of shot noise the Gaussian noise is given by:
\be 
\langle \hat{\delta}^{(1)}(\vk_{l})\hat{\delta}^{(1)*}(\vk_{l}') \rangle = (2\pi)^3 \delta_{\rm D}(\vk_{l}-\vk_{l}')[P_{\rm lin}(k_{l})+N(\vk_l)]
\ee 
with 
\begin{eqnarray}
&&N(\vec{k}_{l})=2A^2(\vk_{l})\vs
&&\times\int \frac{d^3 \vec{k}_{s}}{(2\pi)^3} g^2(\vec{k}_{s},\vk_l-\vec{k}_{s})P_{\rm nl}(k_{s})P_{\rm nl}(|\vk_l-\vk_s|)
\end{eqnarray}
where $P_{\rm nl}$ is the nonlinear power spectrum. Minimizing the noise term we can fix the form of $g$ to be:
\begin{eqnarray}
&&g(\vec{k}_{s},\vec{k}_{s}')
=\frac{f(\vec{k}_{s},\vec{k}_{s}')}{2P_{\rm nl}(k_{s})P_{\rm nl}(k_{s}')}\vs
&=&\frac{F_2(-\vec{k}_s,\vec{k}_s+\vec{k}_s')P_{\rm lin}(k_s)+F_2(-\vec{k}_s',\vec{k}_s+\vec{k}_s')P_{\rm lin}(k_s')}{P_{\rm nl}(k_{s})P_{\rm nl}(k_{s}')}\vs 
\end{eqnarray} 
The noise term reduces simply to $N(\vk_l)=A(\vk_l)$. We find by testing that the value of $N$ is very insensitive to the choice of the lower limit of the integration \ec{a}, since most of the contribution comes from large $k_s$. 

Assuming Gaussian noise, the projected detectability of a $P(k_l)$ measurement using the quadratic estimator can be expressed as:
\be
\frac{1}{\sigma^{2}(k_l)}=\frac{V k_l^2 \Delta k }{(2\pi)^2}\bigg[\frac{P_{\rm lin}(k_l)}{P_{\rm lin}(k_l)+N(k_l) }\bigg]^2 \eql{error},
\ee
where $V$ is the volume of a survey and we compute the detectability for a set of narrow $k_l$-bins each separated by width $\Delta k$.
In Fig.~\rf{SN}, we show the projected errors on the long-wavelength power spectrum using this quadratic estimator in a large survey. The current largest scale published measurement of the three dimensional power spectrum is for scale $0.02 \, h \,\rm Mpc^{-1}$, from \cite{Gil-Marin:2018SDSS}. And our Fig.~\rf{SN} shows that it should be possible to make measurements using our method on scales of $0.002 \, h \,\rm Mpc^{-1}$, which are $\sim$10 times larger. The upper limit of the $\vk_s$ integration in \ec{a} is set to be $0.22\,h\,\rm Mpc^{-1}$. We will see that this choice of the upper limit is reasonable for our current construction at $z=0$. Also notice that $P_{\rm lin}(k_l)$ dominates over $N(k_l)$ in \ec{error} for this upper limit, thus the projected error bars are only slightly wider than the cosmic variance error bars ($N=0$).

\Sfig{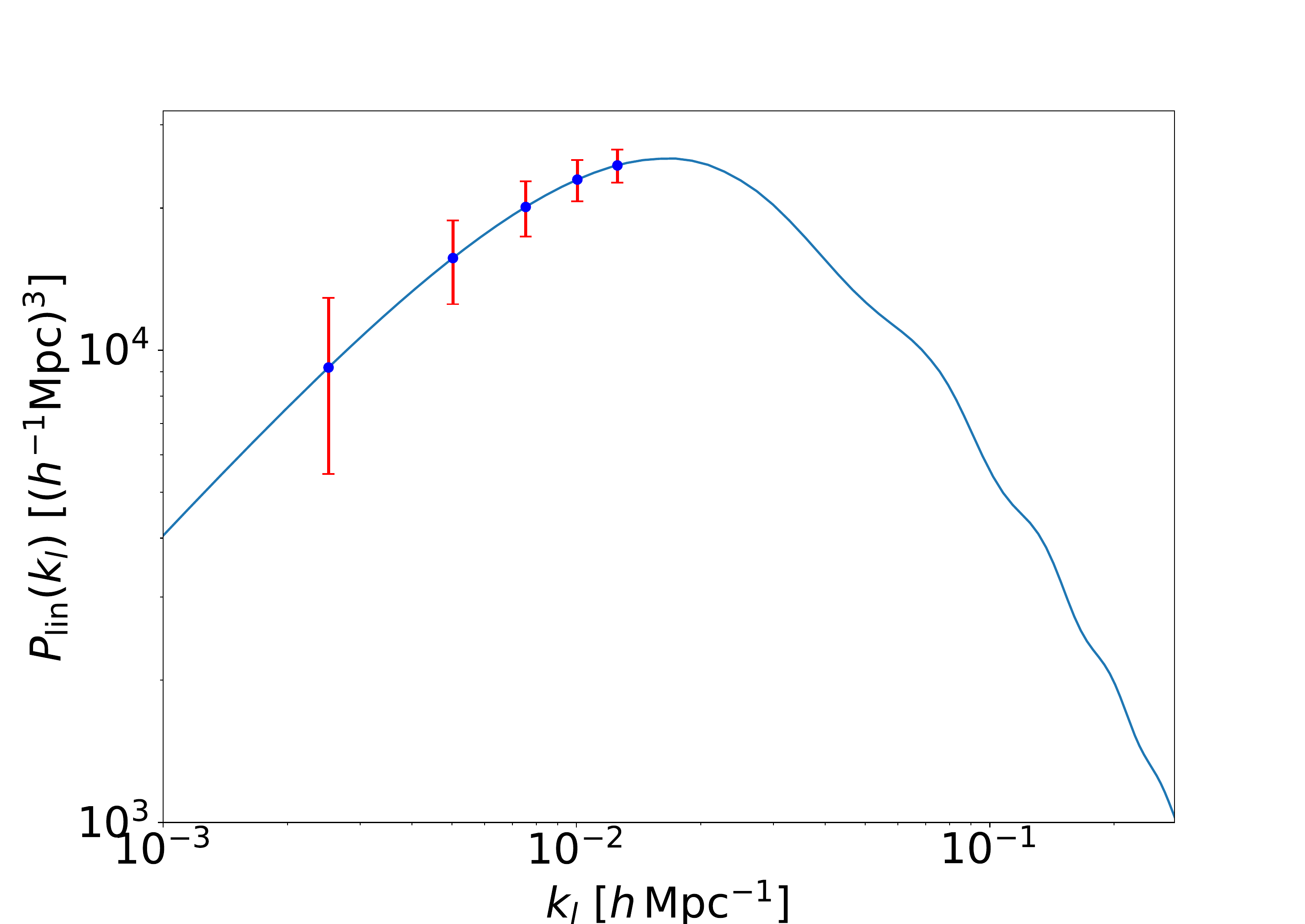}{Long-wavelength power spectrum and its error from \ec{error} which can be expressed as $P(k_l)\sigma(k_l)$. We assume a toy survey of boxsize $L=2.5\, h^{-1}\rm\, Gpc$, thus volume $V=L^3$ and width $\Delta k = 2\pi/L$. Moreover, we set the integration range for $\vk_s$ from $0.03 \,h \,\rm Mpc^{-1}$ to $0.22\,h\,\rm Mpc^{-1}$.}

\section{Demonstration with an N-Body Simulation}\label{sec3}
We test the power of the quadratic estimator using data from a cosmological N-body simulation. We use the $z=0$ snapshot from BigMPDL, one of the MultiDark cosmological simulations \cite{Klypin:2014nov}. The cubical box side length of BigMDPL is $2.5\,h^{-1}\,\rm Gpc$. We use the dark matter particle data to compute the matter density field, leaving the effect of using galaxies or halos to trace the field \cite{Desjacques:2018rev} to  future work.

We use the code nbodykit \cite{Hand:2018nby} to measure the Fourier density modes, and \ec{quadest} to estimate the long wavelength modes  from the measured short wavelength modes. How well the estimator works can be seen from Fig.~\rf{hist}, where we show histograms of the ratio of the estimated mode amplitudes $\hat{\delta}(\vk_l)$ to their true amplitudes $\delta(\vk_l)$ for different values of $\vk_l$. The two panels show the differences between a short wavelength mode cutoff of $k_s=0.22\, h\, \rm Mpc^{-1}$ and  $k_s=0.37\,h\, \rm Mpc^{-1}$. Notice that second-order SPT becomes less accurate as shorter wavelengths are used and will produce a bias of our quadratic estimator. The figure shows that individual mode amplitudes are  unbiased when $k_{s,\rm max}= 0.22\,h\, \rm Mpc^{-1}$, while for $k_{s,\rm max}=0.37\,h\, \rm Mpc^{-1}$, the results are biased (the center of the ratio is $\sim 20 \%$ too high, and the histogram of the polar angle of $\hat{\delta}(\vk_l)/\delta(\vk_l)$ is less peaked at $0$.).
\Sfig{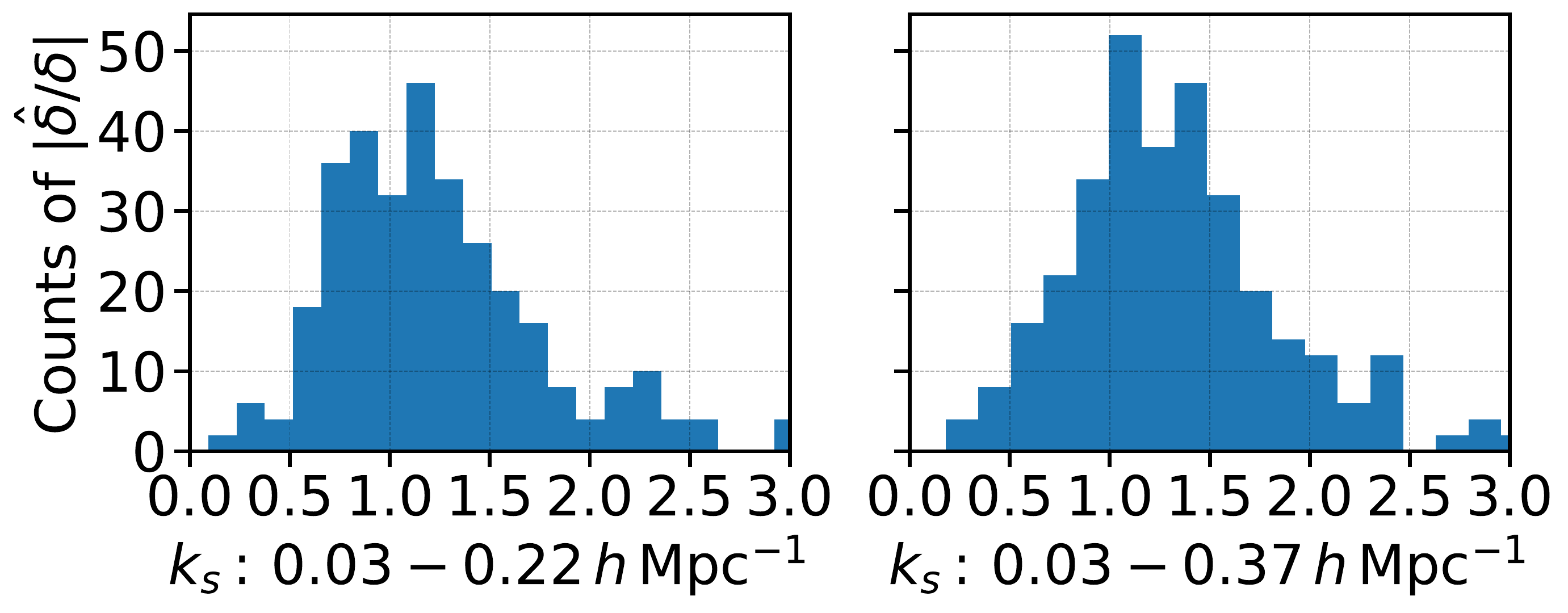}{Comparison of measured and predicted Fourier mode amplitudes: histograms of number counts of $|\hat{\delta}(\vk_l)/\delta(\vk_l)|$ for two different $k_s$ integration ranges, $x$-axis shows the magnitude of $|\hat{\delta}(\vk_l)/\delta(\vk_l)|$.}

Another way of examining the success of the quadratic estimator is to transform the estimated density field back to real space to form $\hat\delta(\vec x)$ and then compare with the actual large scale density field $\delta(\vec x)$ in the simulation. The 7 panels in the top two rows of Fig.~\rf{real} compare these two fields; each panel is a slice of the full simulations. The bottom panel shows the difference between the estimated and true density fields. It is apparent that the differences are much smaller than the overdensities; equivalently the estimator does an excellent job of extracting the large scale density field. 
\section{Conclusion}\label{sec4}
In this paper, we have proposed a new and potentially powerful method to measure long-wavelength modes without having to actually measure large scale structure directly. Similarly to this construction, kSZ velocities \cite{Kosowsky:2009ksz} might also be a good tracer of large-scale modes. We can take the advantage of it's small scale information and potentially get a better constraint of large-scale modes. We will leave this part to future work.

The estimator works well on an N-Body simulation, so applying this estimator to survey data is the logical next step. Among the issues that must be faced when dealing with a spectroscopic galaxy survey are: galaxy bias, redshift space distortions \cite{Kaiser:1987rsd}, and light-cone effects. We do not expect any of these to be show-stoppers, so it is tempting to speculate about the possibilities that will open up with this estimate of the large scale density field. 

First, we can hope to measure 3D clustering on scales larger than the scale entering the horizon at matter-radiation equality without worrying about large-scale systematic effects. General relativistic effects are strongest on large scales (e.g., \cite{Jeong:2012ls}), and these could be detected. There is evidence of large scale anomalies, in the CMB, that could be confronted with maps of large scale structure obtained with this estimator. One physical mechanism that has been proposed as a possible explanation for the deficit in the large-angle CMB temperature correlations is a suppression of primordial power on $\sim$ Gpc scales \cite{Hearin:2011anom}. It would be very useful to verify if this new physics is also present in other probes of large-scale structure. Primordial non-Gaussianity generated by inflation leaves an imprint on the largest scales. There is even the possibility of cross-correlating the large-scale matter field with the CMB itself to extract information about the longest wavelength modes in the universe. Since the current epoch of acceleration is a large-scale, late-time effect, there is the possibility of learning about the mechanism responsible for acceleration. Although, as mentioned above, challenges remain, there is also the possibility of using even smaller wavelength modes in our estimator by going to higher order in perturbation theory.  
\acknowledgements
We thank Adam Solomon, Duncan Campbell, Fabian Schmidt and Lam Hui for resourceful discussions.  This work is supported by U.S.\ Dept.\ of Energy contract DE-SC0019248 and NSF AST-1909193.
The BigMDPL simulation was performed at LRZ Munich within the PRACE project pr86bu. The CosmoSim database (\url{www.cosmosim.org}) providing the file access is a service by the Leibniz-Institute for Astrophysics Potsdam (AIP).

\onecolumngrid
\Sfigg{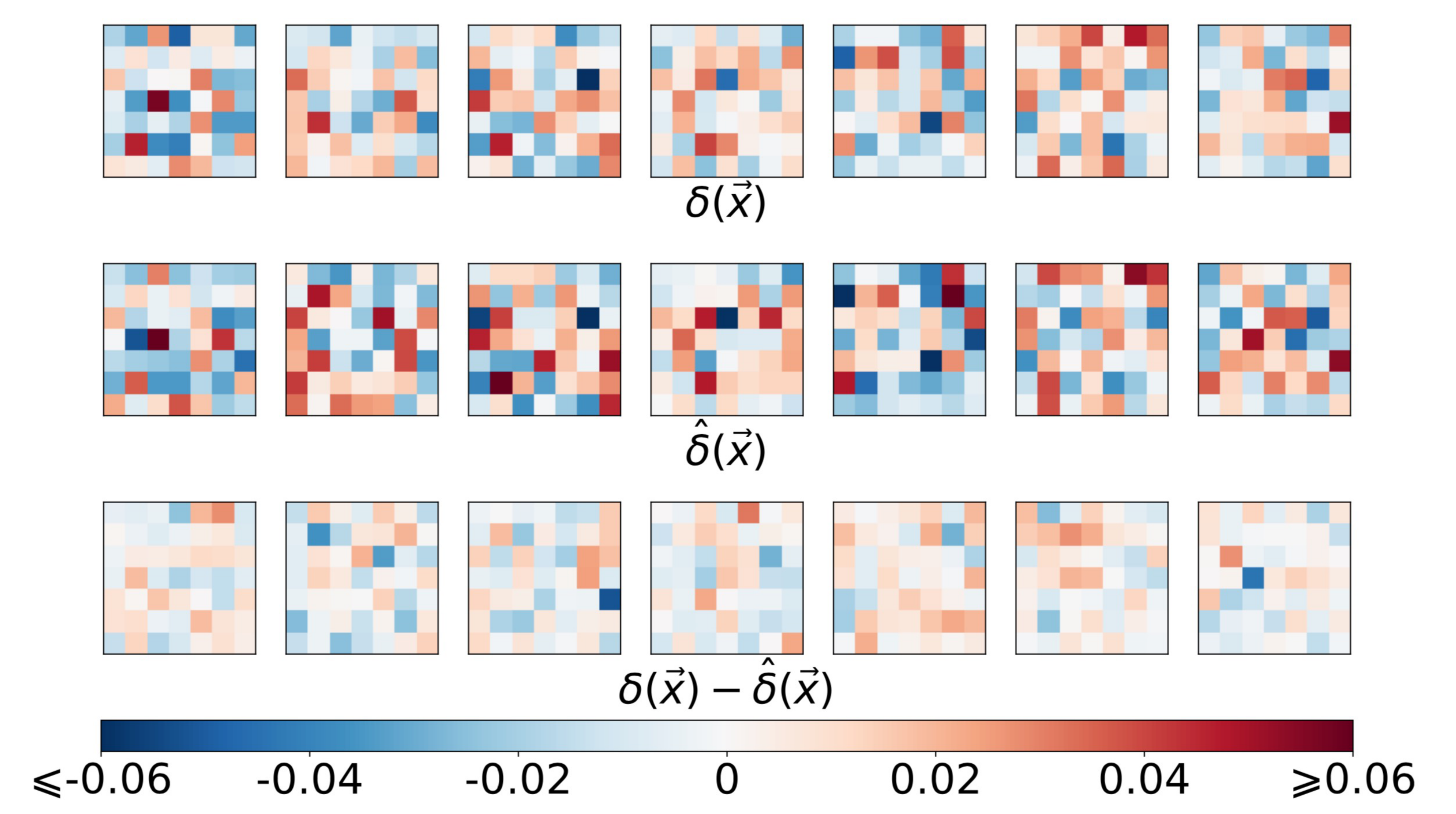}{Comparison of the true density field in the BigMPDL simulation ($\delta(\vec{x})$ computed using the directly measured large-scale modes, top row) and the density field from the quadratic estimator ($\hat{\delta}(\vec{x})$, middle row). The bottom row shows their difference. Each panel represents a slice through the simulation volume,  $2.5 \,h^{-1}\,\rm Gpc$ wide, and one cell ($0.36 \,h^{-1}\,\rm Gpc$) thick.  The upper limit of $\vk_s$ is $0.22 \,h\, \rm Mpc^{-1}$.}
\twocolumngrid

\clearpage

\begin{thebibliography}{34}%
\makeatletter
\providecommand \@ifxundefined [1]{%
 \@ifx{#1\undefined}
}%
\providecommand \@ifnum [1]{%
 \ifnum #1\expandafter \@firstoftwo
 \else \expandafter \@secondoftwo
 \fi
}%
\providecommand \@ifx [1]{%
 \ifx #1\expandafter \@firstoftwo
 \else \expandafter \@secondoftwo
 \fi
}%
\providecommand \natexlab [1]{#1}%
\providecommand \enquote  [1]{``#1''}%
\providecommand \bibnamefont  [1]{#1}%
\providecommand \bibfnamefont [1]{#1}%
\providecommand \citenamefont [1]{#1}%
\providecommand \href@noop [0]{\@secondoftwo}%
\providecommand \href [0]{\begingroup \@sanitize@url \@href}%
\providecommand \@href[1]{\@@startlink{#1}\@@href}%
\providecommand \@@href[1]{\endgroup#1\@@endlink}%
\providecommand \@sanitize@url [0]{\catcode `\\12\catcode `\$12\catcode
  `\&12\catcode `\#12\catcode `\^12\catcode `\_12\catcode `\%12\relax}%
\providecommand \@@startlink[1]{}%
\providecommand \@@endlink[0]{}%
\providecommand \url  [0]{\begingroup\@sanitize@url \@url }%
\providecommand \@url [1]{\endgroup\@href {#1}{\urlprefix }}%
\providecommand \urlprefix  [0]{URL }%
\providecommand \Eprint [0]{\href }%
\providecommand \doibase [0]{http://dx.doi.org/}%
\providecommand \selectlanguage [0]{\@gobble}%
\providecommand \bibinfo  [0]{\@secondoftwo}%
\providecommand \bibfield  [0]{\@secondoftwo}%
\providecommand \translation [1]{[#1]}%
\providecommand \BibitemOpen [0]{}%
\providecommand \bibitemStop [0]{}%
\providecommand \bibitemNoStop [0]{.\EOS\space}%
\providecommand \EOS [0]{\spacefactor3000\relax}%
\providecommand \BibitemShut  [1]{\csname bibitem#1\endcsname}%
\let\auto@bib@innerbib\@empty
\bibitem [{\citenamefont {Collaboration}()}]{LSST:2012ls}%
  \BibitemOpen
  \bibfield  {author} {\bibinfo {author} {\bibfnamefont {L.~D. E.~S.}\
  \bibnamefont {Collaboration}},\ }\href@noop {} {\ }\Eprint
  {http://arxiv.org/abs/1211.0310} {arXiv:1211.0310 [astro-ph.CO]} \BibitemShut
  {NoStop}%
\bibitem [{\citenamefont {Team}()}]{Wfirst:2012jg}%
  \BibitemOpen
  \bibfield  {author} {\bibinfo {author} {\bibfnamefont {W.~S.~D.}\
  \bibnamefont {Team}},\ }\href@noop {} {\ }\Eprint
  {http://arxiv.org/abs/1208.4012} {arXiv:1208.4012 [astro-ph.IM]} \BibitemShut
  {NoStop}%
\bibitem [{\citenamefont {Karkare}(2019)}]{Karkare:2019delen}%
  \BibitemOpen
  \bibfield  {author} {\bibinfo {author} {\bibfnamefont {K.~S.}\ \bibnamefont
  {Karkare}},\ }\href {\doibase 10.1103/PhysRevD.100.043529} {\bibfield
  {journal} {\bibinfo  {journal} {Phys. Rev. D}\ }\textbf {\bibinfo {volume}
  {100}},\ \bibinfo {pages} {043529} (\bibinfo {year} {2019})},\ \Eprint
  {http://arxiv.org/abs/1908.08128} {arXiv:1908.08128 [astro-ph.CO]}
  \BibitemShut {NoStop}%
\bibitem [{\citenamefont {Modi}\ \emph {et~al.}(2019)\citenamefont {Modi},
  \citenamefont {White}, \citenamefont {Slosar},\ and\ \citenamefont
  {Castorina}}]{Modi:2019hydr}%
  \BibitemOpen
  \bibfield  {author} {\bibinfo {author} {\bibfnamefont {C.}~\bibnamefont
  {Modi}}, \bibinfo {author} {\bibfnamefont {M.}~\bibnamefont {White}},
  \bibinfo {author} {\bibfnamefont {A.}~\bibnamefont {Slosar}}, \ and\ \bibinfo
  {author} {\bibfnamefont {E.}~\bibnamefont {Castorina}},\ }\href {\doibase
  10.1088/1475-7516/2019/11/023} {\bibfield  {journal} {\bibinfo  {journal} {J.
  Cosmol. Astropart. P.}\ }\textbf {\bibinfo {volume} {11}},\ \bibinfo {pages}
  {023} (\bibinfo {year} {2019})},\ \Eprint {http://arxiv.org/abs/1907.02330}
  {arXiv:1907.02330 [astro-ph.CO]} \BibitemShut {NoStop}%
\bibitem [{\citenamefont {Masui}\ and\ \citenamefont
  {Pen}(2010)}]{Masui:2010prim}%
  \BibitemOpen
  \bibfield  {author} {\bibinfo {author} {\bibfnamefont {K.~W.}\ \bibnamefont
  {Masui}}\ and\ \bibinfo {author} {\bibfnamefont {U.-L.}\ \bibnamefont
  {Pen}},\ }\href {\doibase 10.1103/PhysRevLett.105.161302} {\bibfield
  {journal} {\bibinfo  {journal} {Phys. Rev. Lett.}\ }\textbf {\bibinfo
  {volume} {105}},\ \bibinfo {pages} {161302} (\bibinfo {year} {2010})},\
  \Eprint {http://arxiv.org/abs/1006.4181} {arXiv:1006.4181 [astro-ph.CO]}
  \BibitemShut {NoStop}%
\bibitem [{\citenamefont {Baldauf}\ \emph {et~al.}(2011)\citenamefont
  {Baldauf}, \citenamefont {Seljak}, \citenamefont {Senatore},\ and\
  \citenamefont {Zaldarriaga}}]{Baldauf:2011fer}%
  \BibitemOpen
  \bibfield  {author} {\bibinfo {author} {\bibfnamefont {T.}~\bibnamefont
  {Baldauf}}, \bibinfo {author} {\bibfnamefont {U.}~\bibnamefont {Seljak}},
  \bibinfo {author} {\bibfnamefont {L.}~\bibnamefont {Senatore}}, \ and\
  \bibinfo {author} {\bibfnamefont {M.}~\bibnamefont {Zaldarriaga}},\ }\href
  {\doibase 10.1088/1475-7516/2011/10/031} {\bibfield  {journal} {\bibinfo
  {journal} {J. Cosmol. Astropart. P.}\ }\textbf {\bibinfo {volume} {10}},\
  \bibinfo {pages} {031} (\bibinfo {year} {2011})},\ \Eprint
  {http://arxiv.org/abs/1106.5507} {arXiv:1106.5507 [astro-ph.CO]} \BibitemShut
  {NoStop}%
\bibitem [{\citenamefont {Jeong}\ and\ \citenamefont
  {Kamionkowski}(2012)}]{Jeong:2012foss}%
  \BibitemOpen
  \bibfield  {author} {\bibinfo {author} {\bibfnamefont {D.}~\bibnamefont
  {Jeong}}\ and\ \bibinfo {author} {\bibfnamefont {M.}~\bibnamefont
  {Kamionkowski}},\ }\href {\doibase 10.1103/PhysRevLett.108.251301} {\bibfield
   {journal} {\bibinfo  {journal} {Phys. Rev. Lett.}\ }\textbf {\bibinfo
  {volume} {108}},\ \bibinfo {pages} {251301} (\bibinfo {year} {2012})},\
  \Eprint {http://arxiv.org/abs/1203.0302} {arXiv:1203.0302 [astro-ph.CO]}
  \BibitemShut {NoStop}%
\bibitem [{\citenamefont {Jasche}\ and\ \citenamefont
  {Wandelt}(2013)}]{Jasche:2013bay}%
  \BibitemOpen
  \bibfield  {author} {\bibinfo {author} {\bibfnamefont {J.}~\bibnamefont
  {Jasche}}\ and\ \bibinfo {author} {\bibfnamefont {B.~D.}\ \bibnamefont
  {Wandelt}},\ }\href {\doibase 10.1093/mnras/stt449} {\bibfield  {journal}
  {\bibinfo  {journal} {Mon. Not. Roy. Astron. Soc.}\ }\textbf {\bibinfo
  {volume} {432}},\ \bibinfo {pages} {894–913} (\bibinfo {year} {2013})},\
  \Eprint {http://arxiv.org/abs/1203.3639} {arXiv:1203.3639 [astro-ph.CO]}
  \BibitemShut {NoStop}%
\bibitem [{\citenamefont {Leclercq}\ \emph {et~al.}(2013)\citenamefont
  {Leclercq}, \citenamefont {Jasche}, \citenamefont {Gil-Marín},\ and\
  \citenamefont {Wandelt}}]{Leclercq:2013remp}%
  \BibitemOpen
  \bibfield  {author} {\bibinfo {author} {\bibfnamefont {F.}~\bibnamefont
  {Leclercq}}, \bibinfo {author} {\bibfnamefont {J.}~\bibnamefont {Jasche}},
  \bibinfo {author} {\bibfnamefont {H.}~\bibnamefont {Gil-Marín}}, \ and\
  \bibinfo {author} {\bibfnamefont {B.~D.}\ \bibnamefont {Wandelt}},\ }\href
  {\doibase 10.1088/1475-7516/2013/11/048} {\bibfield  {journal} {\bibinfo
  {journal} {J. Cosmol. Astropart. P.}\ }\textbf {\bibinfo {volume} {11}},\
  \bibinfo {pages} {048} (\bibinfo {year} {2013})},\ \Eprint
  {http://arxiv.org/abs/1305.4642} {arXiv:1305.4642 [astro-ph.CO]} \BibitemShut
  {NoStop}%
\bibitem [{\citenamefont {Li}\ \emph {et~al.}(2014)\citenamefont {Li},
  \citenamefont {Hu},\ and\ \citenamefont {Takada}}]{Li:2014ssc}%
  \BibitemOpen
  \bibfield  {author} {\bibinfo {author} {\bibfnamefont {Y.}~\bibnamefont
  {Li}}, \bibinfo {author} {\bibfnamefont {W.}~\bibnamefont {Hu}}, \ and\
  \bibinfo {author} {\bibfnamefont {M.}~\bibnamefont {Takada}},\ }\href
  {\doibase 10.1103/PhysRevD.90.103530} {\bibfield  {journal} {\bibinfo
  {journal} {Phys. Rev. D}\ }\textbf {\bibinfo {volume} {90}},\ \bibinfo
  {pages} {103530} (\bibinfo {year} {2014})},\ \Eprint
  {http://arxiv.org/abs/1408.1081} {arXiv:1408.1081 [astro-ph.CO]} \BibitemShut
  {NoStop}%
\bibitem [{\citenamefont {Zhu}\ \emph {et~al.}(2016)\citenamefont {Zhu},
  \citenamefont {Pen}, \citenamefont {Yu}, \citenamefont {Er},\ and\
  \citenamefont {Chen}}]{Zhu:2016tidal}%
  \BibitemOpen
  \bibfield  {author} {\bibinfo {author} {\bibfnamefont {H.-M.}\ \bibnamefont
  {Zhu}}, \bibinfo {author} {\bibfnamefont {U.-L.}\ \bibnamefont {Pen}},
  \bibinfo {author} {\bibfnamefont {Y.}~\bibnamefont {Yu}}, \bibinfo {author}
  {\bibfnamefont {X.}~\bibnamefont {Er}}, \ and\ \bibinfo {author}
  {\bibfnamefont {X.}~\bibnamefont {Chen}},\ }\href {\doibase
  10.1103/PhysRevD.93.103504} {\bibfield  {journal} {\bibinfo  {journal} {Phys.
  Rev. D}\ }\textbf {\bibinfo {volume} {93}},\ \bibinfo {pages} {103504}
  (\bibinfo {year} {2016})},\ \Eprint {http://arxiv.org/abs/1511.04680}
  {arXiv:1511.04680 [astro-ph.CO]} \BibitemShut {NoStop}%
\bibitem [{\citenamefont {Barreira}\ and\ \citenamefont
  {Schimidt}(2017)}]{Barreira:2017res}%
  \BibitemOpen
  \bibfield  {author} {\bibinfo {author} {\bibfnamefont {A.}~\bibnamefont
  {Barreira}}\ and\ \bibinfo {author} {\bibfnamefont {F.}~\bibnamefont
  {Schimidt}},\ }\href {\doibase 10.1088/1475-7516/2017/06/053} {\bibfield
  {journal} {\bibinfo  {journal} {J. Cosmol. Astropart. P.}\ }\textbf {\bibinfo
  {volume} {06}},\ \bibinfo {pages} {03} (\bibinfo {year} {2017})},\ \Eprint
  {http://arxiv.org/abs/1703.09212} {arXiv:1703.09212 [astro-ph.CO]}
  \BibitemShut {NoStop}%
\bibitem [{\citenamefont {Seljak}\ \emph {et~al.}(2017)\citenamefont {Seljak},
  \citenamefont {Aslanyan}, \citenamefont {Feng},\ and\ \citenamefont
  {Modi}}]{Seljak:2017opt}%
  \BibitemOpen
  \bibfield  {author} {\bibinfo {author} {\bibfnamefont {U.}~\bibnamefont
  {Seljak}}, \bibinfo {author} {\bibfnamefont {G.}~\bibnamefont {Aslanyan}},
  \bibinfo {author} {\bibfnamefont {Y.}~\bibnamefont {Feng}}, \ and\ \bibinfo
  {author} {\bibfnamefont {C.}~\bibnamefont {Modi}},\ }\href {\doibase
  10.1088/1475-7516/2017/12/009} {\bibfield  {journal} {\bibinfo  {journal} {J.
  Cosmol. Astropart. P.}\ }\textbf {\bibinfo {volume} {11}},\ \bibinfo {pages}
  {009} (\bibinfo {year} {2017})},\ \Eprint {http://arxiv.org/abs/1706.06645}
  {arXiv:1706.06645 [astro-ph.CO]} \BibitemShut {NoStop}%
\bibitem [{\citenamefont {Akitsu}\ and\ \citenamefont
  {Takada}(2018)}]{Akitsu:2018dist}%
  \BibitemOpen
  \bibfield  {author} {\bibinfo {author} {\bibfnamefont {K.}~\bibnamefont
  {Akitsu}}\ and\ \bibinfo {author} {\bibfnamefont {M.}~\bibnamefont
  {Takada}},\ }\href {\doibase 10.1103/PhysRevD.97.063527} {\bibfield
  {journal} {\bibinfo  {journal} {Phys. Rev. D}\ }\textbf {\bibinfo {volume}
  {97}},\ \bibinfo {pages} {063527} (\bibinfo {year} {2018})},\ \Eprint
  {http://arxiv.org/abs/1711.00012} {arXiv:1711.00012 [astro-ph.CO]}
  \BibitemShut {NoStop}%
\bibitem [{\citenamefont {Foreman}\ \emph {et~al.}(2018)\citenamefont
  {Foreman}, \citenamefont {Meerburg}, \citenamefont {Engelen},\ and\
  \citenamefont {Meyers}}]{Foreman:2018line}%
  \BibitemOpen
  \bibfield  {author} {\bibinfo {author} {\bibfnamefont {S.}~\bibnamefont
  {Foreman}}, \bibinfo {author} {\bibfnamefont {P.~D.}\ \bibnamefont
  {Meerburg}}, \bibinfo {author} {\bibfnamefont {A.~v.}\ \bibnamefont
  {Engelen}}, \ and\ \bibinfo {author} {\bibfnamefont {J.}~\bibnamefont
  {Meyers}},\ }\href {\doibase 10.1088/1475-7516/2018/07/046} {\bibfield
  {journal} {\bibinfo  {journal} {J. Cosmol. Astropart. P.}\ }\textbf {\bibinfo
  {volume} {07}},\ \bibinfo {pages} {046} (\bibinfo {year} {2018})},\ \Eprint
  {http://arxiv.org/abs/1803.04975} {arXiv:1803.04975 [astro-ph.CO]}
  \BibitemShut {NoStop}%
\bibitem [{\citenamefont {Zhu}\ \emph {et~al.}(2018)\citenamefont {Zhu},
  \citenamefont {Pen}, \citenamefont {Yu},\ and\ \citenamefont
  {Chen}}]{Zhu:2018cm}%
  \BibitemOpen
  \bibfield  {author} {\bibinfo {author} {\bibfnamefont {H.-M.}\ \bibnamefont
  {Zhu}}, \bibinfo {author} {\bibfnamefont {U.-L.}\ \bibnamefont {Pen}},
  \bibinfo {author} {\bibfnamefont {Y.}~\bibnamefont {Yu}}, \ and\ \bibinfo
  {author} {\bibfnamefont {X.}~\bibnamefont {Chen}},\ }\href {\doibase
  10.1103/PhysRevD.98.043511} {\bibfield  {journal} {\bibinfo  {journal} {Phys.
  Rev. D}\ }\textbf {\bibinfo {volume} {98}},\ \bibinfo {pages} {043511}
  (\bibinfo {year} {2018})},\ \Eprint {http://arxiv.org/abs/1610.07062}
  {arXiv:1610.07062 [astro-ph.CO]} \BibitemShut {NoStop}%
\bibitem [{\citenamefont {Chiang}\ \emph {et~al.}(2014)\citenamefont {Chiang},
  \citenamefont {Wagner}, \citenamefont {Schmidt},\ and\ \citenamefont
  {Komatsu}}]{Chiang:2014pos}%
  \BibitemOpen
  \bibfield  {author} {\bibinfo {author} {\bibfnamefont {C.-T.}\ \bibnamefont
  {Chiang}}, \bibinfo {author} {\bibfnamefont {C.}~\bibnamefont {Wagner}},
  \bibinfo {author} {\bibfnamefont {F.}~\bibnamefont {Schmidt}}, \ and\
  \bibinfo {author} {\bibfnamefont {E.}~\bibnamefont {Komatsu}},\ }\href
  {\doibase 10.1088/1475-7516/2014/05/048} {\bibfield  {journal} {\bibinfo
  {journal} {J. Cosmol. Astropart. P.}\ }\textbf {\bibinfo {volume} {05}},\
  \bibinfo {pages} {048} (\bibinfo {year} {2014})},\ \Eprint
  {http://arxiv.org/abs/1403.3411} {arXiv:1403.3411 [astro-ph.CO]} \BibitemShut
  {NoStop}%
\bibitem [{\citenamefont {Chiang}\ \emph {et~al.}(2015)\citenamefont {Chiang},
  \citenamefont {Wagner}, \citenamefont {Sánchez}, \citenamefont {Schmidt},\
  and\ \citenamefont {Komatsu}}]{Chiang:2015poss}%
  \BibitemOpen
  \bibfield  {author} {\bibinfo {author} {\bibfnamefont {C.-T.}\ \bibnamefont
  {Chiang}}, \bibinfo {author} {\bibfnamefont {C.}~\bibnamefont {Wagner}},
  \bibinfo {author} {\bibfnamefont {A.~G.}\ \bibnamefont {Sánchez}}, \bibinfo
  {author} {\bibfnamefont {F.}~\bibnamefont {Schmidt}}, \ and\ \bibinfo
  {author} {\bibfnamefont {E.}~\bibnamefont {Komatsu}},\ }\href {\doibase
  10.1088/1475-7516/2015/09/028} {\bibfield  {journal} {\bibinfo  {journal} {J.
  Cosmol. Astropart. P.}\ }\textbf {\bibinfo {volume} {09}},\ \bibinfo {pages}
  {028} (\bibinfo {year} {2015})},\ \Eprint {http://arxiv.org/abs/1504.03322}
  {arXiv:1504.03322 [astro-ph.CO]} \BibitemShut {NoStop}%
\bibitem [{\citenamefont {Goroff}\ \emph {et~al.}(1986)\citenamefont {Goroff},
  \citenamefont {Grinstein}, \citenamefont {Rey},\ and\ \citenamefont
  {Wise}}]{Goroff:1986sts}%
  \BibitemOpen
  \bibfield  {author} {\bibinfo {author} {\bibfnamefont {M.~H.}\ \bibnamefont
  {Goroff}}, \bibinfo {author} {\bibfnamefont {B.}~\bibnamefont {Grinstein}},
  \bibinfo {author} {\bibfnamefont {S.-J.}\ \bibnamefont {Rey}}, \ and\
  \bibinfo {author} {\bibfnamefont {M.~B.}\ \bibnamefont {Wise}},\ }\href
  {\doibase 10.1086/164749} {\bibfield  {journal} {\bibinfo  {journal}
  {Astrophys. J.}\ }\textbf {\bibinfo {volume} {311}},\ \bibinfo {pages} {6}
  (\bibinfo {year} {1986})}\BibitemShut {NoStop}%
\bibitem [{\citenamefont {Makino}\ \emph {et~al.}(1992)\citenamefont {Makino},
  \citenamefont {Misao},\ and\ \citenamefont {Yasushi}}]{Makino:1992fs}%
  \BibitemOpen
  \bibfield  {author} {\bibinfo {author} {\bibfnamefont {N.}~\bibnamefont
  {Makino}}, \bibinfo {author} {\bibfnamefont {S.}~\bibnamefont {Misao}}, \
  and\ \bibinfo {author} {\bibfnamefont {S.}~\bibnamefont {Yasushi}},\ }\href
  {\doibase 10.1103/PhysRevD.46.585} {\bibfield  {journal} {\bibinfo  {journal}
  {Phys. Rev. D}\ }\textbf {\bibinfo {volume} {46}},\ \bibinfo {pages} {585}
  (\bibinfo {year} {1992})}\BibitemShut {NoStop}%
\bibitem [{\citenamefont {Jain}\ and\ \citenamefont
  {Bertschinger}(1994)}]{Jain:1994sop}%
  \BibitemOpen
  \bibfield  {author} {\bibinfo {author} {\bibfnamefont {B.}~\bibnamefont
  {Jain}}\ and\ \bibinfo {author} {\bibfnamefont {E.}~\bibnamefont
  {Bertschinger}},\ }\href {\doibase 10.1086/174502} {\bibfield  {journal}
  {\bibinfo  {journal} {Astrophys. J.}\ }\textbf {\bibinfo {volume} {431}},\
  \bibinfo {pages} {495} (\bibinfo {year} {1994})},\ \Eprint
  {http://arxiv.org/abs/astro-ph/9311070} {arXiv:astro-ph/9311070 [astro-ph]}
  \BibitemShut {NoStop}%
\bibitem [{\citenamefont {Hu}(2001)}]{Hu:2001dt}%
  \BibitemOpen
  \bibfield  {author} {\bibinfo {author} {\bibfnamefont {W.}~\bibnamefont
  {Hu}},\ }\href {\doibase 10.1086/323253} {\bibfield  {journal} {\bibinfo
  {journal} {Astrophys. J.}\ }\textbf {\bibinfo {volume} {557}},\ \bibinfo
  {pages} {L79–L83} (\bibinfo {year} {2001})},\ \Eprint
  {http://arxiv.org/abs/astro-ph/0105424} {arXiv:astro-ph/0105424 [astro-ph]}
  \BibitemShut {NoStop}%
\bibitem [{\citenamefont {Hu}\ and\ \citenamefont {Okamoto}(2002)}]{Hu:2002mr}%
  \BibitemOpen
  \bibfield  {author} {\bibinfo {author} {\bibfnamefont {W.}~\bibnamefont
  {Hu}}\ and\ \bibinfo {author} {\bibfnamefont {T.}~\bibnamefont {Okamoto}},\
  }\href {\doibase 10.1086/341110} {\bibfield  {journal} {\bibinfo  {journal}
  {Astrophys. J.}\ }\textbf {\bibinfo {volume} {574}},\ \bibinfo {pages} {566}
  (\bibinfo {year} {2002})},\ \Eprint {http://arxiv.org/abs/astro-ph/0111606}
  {arXiv:astro-ph/0111606 [astro-ph]} \BibitemShut {NoStop}%
\bibitem [{\citenamefont {{Planck Collaboration}}(2014)}]{Planck:2014cos}%
  \BibitemOpen
  \bibfield  {author} {\bibinfo {author} {\bibnamefont {{Planck
  Collaboration}}},\ }\href {\doibase 10.1051/0004-6361/201321591} {\bibfield
  {journal} {\bibinfo  {journal} {Astron. Astrophys.}\ }\textbf {\bibinfo
  {volume} {571}},\ \bibinfo {pages} {66} (\bibinfo {year} {2014})},\ \Eprint
  {http://arxiv.org/abs/1303.5076} {arXiv:1303.5076 [astro-ph.CO]} \BibitemShut
  {NoStop}%
\bibitem [{\citenamefont {Bernardeau}\ \emph {et~al.}(2012)\citenamefont
  {Bernardeau}, \citenamefont {Colombi}, \citenamefont {Gaztañaga},\ and\
  \citenamefont {Scoccimarro}}]{Bernardeau:2002rev}%
  \BibitemOpen
  \bibfield  {author} {\bibinfo {author} {\bibfnamefont {F.}~\bibnamefont
  {Bernardeau}}, \bibinfo {author} {\bibfnamefont {S.}~\bibnamefont {Colombi}},
  \bibinfo {author} {\bibfnamefont {E.}~\bibnamefont {Gaztañaga}}, \ and\
  \bibinfo {author} {\bibnamefont {Scoccimarro}},\ }\href {\doibase
  10.1016/S0370-1573(02)00135-7} {\bibfield  {journal} {\bibinfo  {journal}
  {Phys. Rept.}\ }\textbf {\bibinfo {volume} {367}},\ \bibinfo {pages} {1}
  (\bibinfo {year} {2012})},\ \Eprint {http://arxiv.org/abs/astro-ph/0112551}
  {arXiv:astro-ph/0112551 [astro-ph]} \BibitemShut {NoStop}%
\bibitem [{\citenamefont {Takahashi}(2008)}]{Takahashi:2008to}%
  \BibitemOpen
  \bibfield  {author} {\bibinfo {author} {\bibfnamefont {R.}~\bibnamefont
  {Takahashi}},\ }\href {\doibase 10.1143/PTP.120.549} {\bibfield  {journal}
  {\bibinfo  {journal} {Prog. Theor. Phys.}\ }\textbf {\bibinfo {volume}
  {120}},\ \bibinfo {pages} {549–559} (\bibinfo {year} {2008})},\ \Eprint
  {http://arxiv.org/abs/0806.1437} {arXiv:0806.1437 [astro-ph.CO]} \BibitemShut
  {NoStop}%
\bibitem [{\citenamefont {Gil-Marín}\ \emph {et~al.}(2018)\citenamefont
  {Gil-Marín} \emph {et~al.}}]{Gil-Marin:2018SDSS}%
  \BibitemOpen
  \bibfield  {author} {\bibinfo {author} {\bibfnamefont {H.}~\bibnamefont
  {Gil-Marín}} \emph {et~al.},\ }\href {\doibase 10.1093/mnras/stt449}
  {\bibfield  {journal} {\bibinfo  {journal} {Mon. Not. Roy. Astron. Soc.}\
  }\textbf {\bibinfo {volume} {477}},\ \bibinfo {pages} {1604} (\bibinfo {year}
  {2018})},\ \Eprint {http://arxiv.org/abs/1801.02689} {arXiv:1801.02689
  [astro-ph.CO]} \BibitemShut {NoStop}%
\bibitem [{\citenamefont {Klypin}\ \emph {et~al.}(2014)\citenamefont {Klypin},
  \citenamefont {Yepes}, \citenamefont {Gottlober}, \citenamefont {Prada},\
  and\ \citenamefont {Hess}}]{Klypin:2014nov}%
  \BibitemOpen
  \bibfield  {author} {\bibinfo {author} {\bibfnamefont {A.}~\bibnamefont
  {Klypin}}, \bibinfo {author} {\bibfnamefont {G.}~\bibnamefont {Yepes}},
  \bibinfo {author} {\bibfnamefont {S.}~\bibnamefont {Gottlober}}, \bibinfo
  {author} {\bibfnamefont {F.}~\bibnamefont {Prada}}, \ and\ \bibinfo {author}
  {\bibfnamefont {S.}~\bibnamefont {Hess}},\ }\href {\doibase
  10.1093/mnras/stw248} {\bibfield  {journal} {\bibinfo  {journal} {Mon. Not.
  Roy. Astron. Soc.}\ }\textbf {\bibinfo {volume} {457}},\ \bibinfo {pages}
  {4340} (\bibinfo {year} {2014})},\ \Eprint {http://arxiv.org/abs/1411.4001}
  {arXiv:1411.4001 [astro-ph.CO]} \BibitemShut {NoStop}%
\bibitem [{\citenamefont {Desjacques}\ \emph {et~al.}(2018)\citenamefont
  {Desjacques}, \citenamefont {Jeong},\ and\ \citenamefont
  {Schmidt}}]{Desjacques:2018rev}%
  \BibitemOpen
  \bibfield  {author} {\bibinfo {author} {\bibfnamefont {V.}~\bibnamefont
  {Desjacques}}, \bibinfo {author} {\bibfnamefont {D.}~\bibnamefont {Jeong}}, \
  and\ \bibinfo {author} {\bibfnamefont {F.}~\bibnamefont {Schmidt}},\ }\href
  {\doibase 10.1016/j.physrep.2017.12.002} {\bibfield  {journal} {\bibinfo
  {journal} {Phys. Rept.}\ }\textbf {\bibinfo {volume} {733}},\ \bibinfo
  {pages} {1} (\bibinfo {year} {2018})},\ \Eprint
  {http://arxiv.org/abs/1611.09787} {arXiv:1611.09787 [astro-ph.CO]}
  \BibitemShut {NoStop}%
\bibitem [{\citenamefont {Hand}\ \emph {et~al.}(2018)\citenamefont {Hand},
  \citenamefont {Feng}, \citenamefont {Beutler}, \citenamefont {Li},
  \citenamefont {Modi}, \citenamefont {Seljak},\ and\ \citenamefont
  {Slepian}}]{Hand:2018nby}%
  \BibitemOpen
  \bibfield  {author} {\bibinfo {author} {\bibfnamefont {N.}~\bibnamefont
  {Hand}}, \bibinfo {author} {\bibfnamefont {Y.}~\bibnamefont {Feng}}, \bibinfo
  {author} {\bibfnamefont {F.}~\bibnamefont {Beutler}}, \bibinfo {author}
  {\bibfnamefont {Y.}~\bibnamefont {Li}}, \bibinfo {author} {\bibfnamefont
  {C.}~\bibnamefont {Modi}}, \bibinfo {author} {\bibfnamefont {U.}~\bibnamefont
  {Seljak}}, \ and\ \bibinfo {author} {\bibfnamefont {Z.}~\bibnamefont
  {Slepian}},\ }\href {\doibase 10.3847/1538-3881/aadae0} {\bibfield  {journal}
  {\bibinfo  {journal} {Astrophys. J.}\ }\textbf {\bibinfo {volume} {156}},\
  \bibinfo {pages} {160} (\bibinfo {year} {2018})},\ \Eprint
  {http://arxiv.org/abs/1712.05834} {arXiv:1712.05834 [astro-ph.CO]}
  \BibitemShut {NoStop}%
\bibitem [{\citenamefont {Kosowsky}\ and\ \citenamefont
  {Bhattacharya}(2009)}]{Kosowsky:2009ksz}%
  \BibitemOpen
  \bibfield  {author} {\bibinfo {author} {\bibfnamefont {A.}~\bibnamefont
  {Kosowsky}}\ and\ \bibinfo {author} {\bibfnamefont {S.}~\bibnamefont
  {Bhattacharya}},\ }\href {\doibase 10.1103/PhysRevD.80.062003} {\bibfield
  {journal} {\bibinfo  {journal} {Phys. Rev. D}\ }\textbf {\bibinfo {volume}
  {80}},\ \bibinfo {pages} {062003} (\bibinfo {year} {2009})},\ \Eprint
  {http://arxiv.org/abs/0907.4202} {arXiv:0907.4202 [astro-ph.CO]} \BibitemShut
  {NoStop}%
\bibitem [{\citenamefont {Kaiser}(1987)}]{Kaiser:1987rsd}%
  \BibitemOpen
  \bibfield  {author} {\bibinfo {author} {\bibfnamefont {N.}~\bibnamefont
  {Kaiser}},\ }\href {\doibase 10.1093/mnras/227.1.1} {\bibfield  {journal}
  {\bibinfo  {journal} {Mon. Not. Roy. Astron. Soc.}\ }\textbf {\bibinfo
  {volume} {227}},\ \bibinfo {pages} {1} (\bibinfo {year} {1987})}\BibitemShut
  {NoStop}%
\bibitem [{\citenamefont {Jeong}\ \emph {et~al.}(2012)\citenamefont {Jeong},
  \citenamefont {Schmidt},\ and\ \citenamefont {Hirata}}]{Jeong:2012ls}%
  \BibitemOpen
  \bibfield  {author} {\bibinfo {author} {\bibfnamefont {D.}~\bibnamefont
  {Jeong}}, \bibinfo {author} {\bibfnamefont {F.}~\bibnamefont {Schmidt}}, \
  and\ \bibinfo {author} {\bibfnamefont {C.~M.}\ \bibnamefont {Hirata}},\
  }\href {\doibase 10.1103/PhysRevD.85.023504} {\bibfield  {journal} {\bibinfo
  {journal} {Phys. Rev. D}\ }\textbf {\bibinfo {volume} {85}},\ \bibinfo
  {pages} {023504} (\bibinfo {year} {2012})},\ \Eprint
  {http://arxiv.org/abs/1107.5427} {arXiv:1107.5427 [astro-ph.CO]} \BibitemShut
  {NoStop}%
\bibitem [{\citenamefont {Hearin}\ \emph {et~al.}(2019)\citenamefont {Hearin},
  \citenamefont {Gibelyou},\ and\ \citenamefont {Zentner}}]{Hearin:2011anom}%
  \BibitemOpen
  \bibfield  {author} {\bibinfo {author} {\bibfnamefont {A.~P.}\ \bibnamefont
  {Hearin}}, \bibinfo {author} {\bibfnamefont {C.}~\bibnamefont {Gibelyou}}, \
  and\ \bibinfo {author} {\bibfnamefont {A.~R.}\ \bibnamefont {Zentner}},\
  }\href {\doibase 10.1088/1475-7516/2011/10/012} {\bibfield  {journal}
  {\bibinfo  {journal} {J. Cosmol. Astropart. P.}\ }\textbf {\bibinfo {volume}
  {10}},\ \bibinfo {pages} {012} (\bibinfo {year} {2019})},\ \Eprint
  {http://arxiv.org/abs/1108.2269} {arXiv:1108.2269 [astro-ph.CO]} \BibitemShut
  {NoStop}%
\end{thebibliography}
%

\end{document}